\documentclass[twocolumn,aps,prl,superscriptaddress]{revtex4}%
\usepackage{color}
\usepackage{amsmath}
\usepackage{amsfonts}
\usepackage{amssymb}
\usepackage{graphicx}%
\providecommand{\U}[1]{\protect\rule{.1in}{.1in}}
\begin{document}
\title{Measurement and Particle Statistics in the Szilard Engine}
\author{Martin Plesch}
\affiliation{Faculty of Informatics, Masaryk University, Brno, Czech Republic}
\affiliation{Institute of Physics, Slovak Academy of Sciences, Bratislava, Slovakia}
\author{Oscar Dahlsten}
\affiliation{Department of Physics, University of Oxford, Clarendon Laboratory, Oxford, OX1
3PU, UK}
\affiliation{Center for Quantum Technology, National University of Singapore, Singapore}
\author{John Goold}
\affiliation{Department of Physics, University of Oxford,
Clarendon Laboratory, Oxford, OX1 3PU, UK}
\author{Vlatko Vedral}
\affiliation{Department of Physics, University of Oxford, Clarendon Laboratory, Oxford, OX1
3PU, UK}
\affiliation{Center for Quantum Technology, National University of Singapore, Singapore}

\begin{abstract}
A Szilard Engine is a hypothetical device which is able to extract
work from a single thermal reservoir by measuring the position of
particles within the engine. We derive the amount of work that can
be extracted from such a device in the low temperature limit.
Interestingly,  we show this work is determined by the information
gain of the initial measurement rather than by the number and type
of particles which constitute the working substance. Our work
provides another clear connection between information gain and
extractable work in thermodynamical processes.
\end{abstract}
\maketitle

%==============================================================================

%==============================================================================

The laws of thermodynamics are well known for their robustness, namely the
fact that they survived both physics revolutions of the 20th century. Neither
quantum physics nor relativity have modified our confidence in the fact that
the overall energy of a closed system should be conserved, while its entropy
tends to the maximum value with time. At low temperatures, however, quantum
systems obey completely different statistics to classical ones and one could
imagine that bosonic and fermionic statistics, while not breaking the laws of
thermodynamics, might still allow us to obtain more work from
indistinguishable particles than from the same number of distinguishable ones.
This has indeed recently been reported in \cite{KSLU}.

In this Letter we show that the performance of a work cycle with
indistinguishable particle depends on the information one has and is capable
of obtaining about them. The type of particles doing the work, be they
classical, quantum, distinguishable or indistinguishable, is only of secondary
importance. Our work clarifies why the basics of thermodynamics are
independent of particle statistics, a fact that makes thermodynamical laws all
the more remarkable. Furthermore, it emphasises that the second law ought to
most appropriately be phrased in terms of a trade-off between information
gained and work done.

In this letter we use the concept of the Szilard engine (SZE)
\cite{Szilard, Szilarbook} - a hypothetical device which consists
of a cylinder, serving as a heat reservoir at temperature $T$,
containing a single gas particle. The engine works by dividing the
cylinder into two halves by an impenetrable barrier and measuring
the position of the gas particle. Following the measurement, work
can be extracted from this device by exploiting the pressure
created by the particle on the barrier. The amount of work is
given as $kT\ln(2)$ with $k$ being the Boltzmann constant. As was
identified by Bennett~\cite{Bennett82}, the validity of the second
thermodynamical law is saved by the fact that the information
about the position of the particle before the work has been done,
or, equivalently, about the position of the barrier after the work
has been done, has to be stored somewhere and erased by
"resetting" the engine, i.e. removing the barrier and moving it to
the middle again.

% Jauch and Baron~\cite{JB} questioned the validity of the thermodynamical
% approach for a one-molecule gas, as well as the work needed to insert the
% barrier. Zurek \cite{Zurek1} showed that a quantum approach can remove these
% doubts by exploiting the intrinsic quantum uncertainty.
 A natural extension of the original SZE is to include more particles into the engine. In e.g.~\cite{Zurek2, Bennett82, LT} it was shown that correlations have a positive impact on the amount of work that can be extracted
from the engine, quantum correlations having a particularly strong
impact. A related important type of multi-particle effect is of
course particle statistics associated with identical particles. In
a recent paper Kim {\it et. al.} \cite{KSLU} performed a full
quantum analysis of a SZE with more than one particle, bringing
the role of particle statistics into focus.

In this Letter we clarify the role of particle statistics in a
quantum SZE by placing a particular emphasis on the initial
measurement. We consider the low temperature limit as we are
interested in quantum effects and these are most strongly
manifested in this regime. We show that the extractable work is
directly connected with the measurement performed on the system.
Crucially, for a measurement with $M$ outcomes, the work is
bounded from above by $W=kT\ln(M)$ and reaches this bound only for
measurements with equiprobable outcomes. In other words, we find
that the extractable work is determined by the information gained
during the initial measurement, regardless of whether the working
medium consists of distinguishable, bosonic or fermionic
particles.
%By
%breaking the process into parts and
%calculating how much work is extracted in each step we determine
%how exactly things conspire to remove the dependence on the
%working medium in this way.
By performing a detailed analysis we show that the work costs of
the different steps do depend on the working medium but conspire
to remove this dependence when combined.

{\em \bf Szilard Engine.---} Consider $N$ particles (bosons,
fermions or distinguishable particles) placed in a cylinder kept
at a constant temperature $T$ via the ongoing interaction between
the particles and the walls of the cylinder. This interaction is
considered to be fast in the sense that the time scale of
thermalization is much smaller than any other time scale used. In
the first part of the Letter we will also assume the energy levels
of the particles in the cylinder, as well as in its parts after
inserting the barrier (piston), to be non-degenerate. However this
does not exclude the energy levels on the respective side of the
piston having the same energies.

The original quantum SZE is used to extract work from a
temperature reservoir via the following steps:

1. A piston (modelled by a sufficiently narrow and high potential
barrier) is inserted into the cylinder to separate it into two
disconnected regions, preventing any tunneling within the relevant
time-scales.   The work cost of this step shall be called $W1$.

2. A measurement is performed to obtain the number of particles on one side of
the cylinder. This measurement might be full, i.e. providing the exact number
of particles (or position in the case of distinguishable particles) or
partial. The state of particles in the cylinder, after the wall was inserted
and before the measurement is performed, is a mixture of possible states
obtained via the measurement rather then their coherent superposition. This is
due to the fact that interactions with the walls of the cylinder will not only
fix the temperature of the gas during the process but also perform a
measurement in the number basis in each part of the cylinder, once the barrier
is introduced into the cylinder.

3. Depending on the result of the measurement the piston will be
allowed to move quasi-statically to a position where the side-ways
force acting on the piston will be zero.

4. The piston is then quasi-statically removed by decreasing the
strength of the potential.

In the {\it classical} case there is no need to invest work into
the engine except during the erasure procedure - insertion of the
piston, its removal as well as the measurement are considered to
be "for free". The only stage of extraction of energy is the
movement of the piston resulting from the unequal pressure on
either side. In contrast, in the the {\it quantum} case one
inevitably has to invest energy to create the barrier (step 1.);
this can however be recovered during the movement and removal of
the barrier (steps 3 and 4). In this picture the stages of
movement and removal of the barrier need not anymore be considered
as independent actions. The extraction of work can be equally well
done by exploiting the force towards the side of the barrier as
well as acting on the top of the barrier. Therefore throughout the
manuscript we will only work with three phases of the engine -
insertion, measurement and movement/removal phase. The combined
work-cost of steps 3. and 4. shall be called $W2$.

These four processes represent a closed cycle up to the point
where the result of the measurement in step $2$ is still stored
somewhere. For a full restoration of the original setting one has
to erase this information~\cite{Bennett82}.

%In this picture the stages of movement and removal of the barrier need not anymore to be considered as independent actions. The extraction of work can be equally well done by exploiting of the force towards the side of the barrier as well as acting on the top of the barrier. Therefore throughout the manuscript we will only work with three phases of the engine - insertion, measurement and movement/removal phase.

%==============================================================================

{\em \bf Low temperature limit.---}In the first part of this
Letter we will consider the system to be non-degenerate and assume
the low temperature limit. Here, the quantum effects of the engine
are expected to manifest themselves in the strongest way - for
high temperatures all particles became essentially distinguishable
as each can be labelled by its energy. In what follows we shall
denote the energy levels of the particles in the cylinder without
the barrier as $E_{i}$. We consider the condition $kT\ll\Delta E,$
where $\Delta E$ represents the difference between any two energy
levels that could be potentially occupied - for bosons and
distinguishable particles it is just the difference between two
lowest energy levels of the system and for fermions $\Delta E$
represents the difference between the Fermi energy and the nearest
higher energy level. Under these assumptions the partition
function of any system consisting of $N$ bosons or distinguishable
particles will be $Z_{b}=\exp\left(  -NE_{1}/ kT\right)  $ with
$E_{1}$ being the lowest energy of the system. For fermions in a
non-degenerate system the partition function will have the form
$Z_{f}=\exp\left(  -\sum_{i=1}^{N}E_{i}/kT\right).$

%==============================================================================

We will now calculate the work that can be extracted from the closed cycle of
the SZE. This can be defined as $W=kT\ln\frac{Z_{B}}{Z_{A}}$ with $Z_{B}$
being the partition function of the final state and $Z_{A}$ being the
partition function of the initial state.

{\em \bf Insertion of the barrier.---}For \textit{bosons} there
are two distinct possibilities for this step. Let us denote the
energy levels on the left(right) of the barrier as $E_{i}^{l(R)}$.
The first one corresponds to the case where we do not insert the
barrier in the middle of the system $E_{1}%
^{l}>E_{1}^{r}$.  Quantities for this case will be labeled by
index $n$ so the final partition function will be
$Z_{b}^{n}=\exp\left(  -NE_{1}^{r}/kT\right) ,$ which physically
means that all particles condensed in the lower potential well.
This will result in a final work $W1_{b}^{n}=N\left(
E_{1}-E_{1}^{r}\right)  .$ Irrespective of the exact form of the
potential defining the cylinder one can expect $E_{1}^{r}>E_{1}$,
as the "living space" of the particles has decreased and thus the
work performed is negative. The second possibility corresponds to
the case when the barrier is introduced in the middle of the
system i.e. $E_{1}^{l}=E_{1}^{r}$. In this case quantities will be
indexed by $d$ and the precision is expected to be such that
$E_{1}^{l}-E_{1}^{r}\ll kT$ holds. Under such an assumption the
partition sum will be $Z_{b}^{d}=(N+1)\exp\left(
-NE_{1}^{r}/kT\right)  $ and the work
$W1_{b}^{d}=kT\ln(N+1)+W1_{b}^{n}$.

For \textit{distinguishable particles} the situation is quite similar to
the above bosonic case. Again there is no limit on the number of particles occupying the lowest
energy level. The only difference is for the degenerate case, where the
particles can choose their positions without change of the total energy of the
system. Here the number of possible configurations will be $2^{N}$ and the
partition function will be $Z_{d}^{d}=2^{N}\exp\left(  -NE_{1}^{r}/kT\right)
$. Here the extractable work $W1_{d}^{d}=N\left(  kT\ln2+E_{1}-E_{1}^{r}\right)
$, simply corresponding to the case of joining $N$ independent SZEs.

For \textit{fermions} the situation is distinctly different. We define $j$
as the level for which following equality holds:%
\begin{equation}
E_{N-j}^{r}\leq E_{j}^{l}\leq E_{N-j+1}^{r}. \label{E_fermions}%
\end{equation}
First let us examine the case where there is sharp inequality on the right hand side of
(\ref{E_fermions}). The final partition function will be $Z_{f}%
^{n}=\exp\left(  -\left(\sum_{i=1}^{j}E_{i}^{l}+\sum_{i=1}^{N-j}E_{i}^{r}\right)/kT\right)$ and the extractable work will be $W1_{f}^{n}=\sum_{i=1}%
^{N}E_{i}-\left(
\sum_{i=1}^{j}E_{i}^{l}+\sum_{i=1}^{N-j}E_{i}^{r}\right) .$ In
contrast, for equality on the right hand side of
(\ref{E_fermions}) we get a degeneracy in the energy levels of the
whole system when the fermion at the Fermi level can freely choose
either side of the cylinder and the partition
function becomes $Z_{f}^{d}=2\exp\left(  -\left(\sum_{i=1}^{j}E_{i}%
^{l}+\sum_{i=1}^{N-j}E_{i}^{r}\right)/kT \right)  $ and the extractable work will be
$W1_{f}^{d}=kT\ln(2)+W1_{f}^{n}.$ It is worth to mention that in contrast to
the bosonic case there are potentially many (in the order of $N$)
possibilities for choosing the position of the barrier to reach the equalized
position, but insertion into the middle of the container will cost less work
only for $N$ odd.

{\em \bf Measurement.---} In the second step a measurement on the
system is performed.  This is only non-trivial if the energy
levels of the final system (after inserting the barrier) are
degenerate (with partition functions and works labelled by $d$) as
in the other case it is just a single-outcome measurement
confirming that all bosons/distinguishable particles are in the
deeper well (larger part of the piston) or the fermions are
distributed within the piston in a way expectable by the
distribution of energy levels.

In the non-trivial bosonic case, the full measurement will have $N+1$ possible
outcomes counting the number of particles on the left hand side. In the
fermionic case the measurement will be binary and for distinguishable
particles the number of possible outcomes will be $2^{N}$ (specifying the position
of every single particle).

{\em \bf Movement and removal of piston.---} In the {\it
classical} case one would extract the work by simply moving the
piston to its equilibrium position. All extractable work would be
extracted within this phase for an infinitely narrow piston. On
the other hand, if the piston was removed from a different
position than the equilibrium one, part of the potential work
would just dissipate due to mixing of gases with different
pressures. In the {\it quantum} case the situation is much more
subtle. Here the extraction of work is not straightforward to
define physically in connection with possible storages of energy
such as excited states of atoms. However, if we stick to the
standard definition of the generalized force as $F=\frac{\partial
E}{\partial\lambda}$ with $\lambda$ being a parameter of the
barrier (e.g. its position during the movement phase or height
during the removal phase) and accept that any such force can be
utilized to perform work, we can calculate the extractable work
from the partition functions without making specific assumptions
on the process itself.

For both bosons and distinguishable particles the initial partition sum of the
system is $Z_{b}^{A}=\exp\left(  -NE_{1}^{r}/kT\right)  $ and final is
$Z_{b}$. Therefore we get as the extractable work $W2_{b}=N\left(  E_{1}%
^{r}-E_{1}\right)  .$ For fermions the partition sum is
$Z_{f}^{A}=\exp\left(
-\left(\sum_{i=1}^{j}E_{i}^{l}+\sum_{i=1}^{N-j}E_{i}^{r}\right)/kT\right)
$ and
the final one $Z_{f}$; the resulting work is $W2_{f}=\sum_{i=1}^{j}E_{i}%
^{l}+\sum_{i=1}^{N-j}E_{i}^{r}-\sum_{i=1}^{N}E_{i}.$

{\em \bf Summing up.---}The total work gained (or paid) during the
relevant parts of the cycle is then $W_{b}=W_{d}=W_{f}=0$ for
non-degenerate cases with trivial measurements. The degenerate
cases are much more interesting. In the case of distinguishable
particles $W_{d}=NkT\ln(2),$ corresponding to $N$ independent
SZEs, for bosons we get $W_{b}=kT\ln(N+1)$ and for fermions
$W_{f}=kT\ln(2).$ All these results (even for non-degenerate
cases) are connected by a unifying formula of the form
\begin{equation}
W=kT\ln(M) \label{NCG_work}%
\end{equation}
with $M$ being the number of possible measurement outcomes, each
of them occurring with equal probability. The work corresponds to
the energy needed to erase a memory able to store the result of
the measurement.

We note that S. W. Kim et. al.~\cite{KSLU} calculate the work
yield during the removal of the barrier differently. They firstly
lower the barrier to the point where tunnelling is practically
unrestricted. They view any energy that could potentially be
extracted here as lost. In the second phase where the barrier is
removed and the "living space" of the particles is enlarged they
do however extract work. With this approach the net energy gained
from the system in the low temperature limit is however always
negative if the number of particles $N>2$. Work could only be
gained if all particles are found on one of the sides of the
cylinder, resulting in extractable work in the order of $kT$ with
probability diminishing with large $N$. On the other hand, all
other measurement results would lead to a loss of energy in the
order of $\Delta E$ via tunnelling
 during the removal of the barrier phase.

%Amount of the energy lost during the phase of partial lowering of
%the barrier is depending on the actual position of the barrier. If
%the barrier is removed from one side of the cylinder, no
%tunnelling occurs resulting in no energy loss. If the barrier is
%removed exactly from the middle of the cylinder, tunnelling is
%possible only between states with equal energy, resulting into
%loss of energy scaled by $kT$. In all other cases energy is lost
%via tunnelling of particles from the squeezed side towards the
%expanded side, jumping from higher towards lower energy levels
%scaling with $\Delta E$ (being the difference between energy
%levels on different sides of the cylinder). With this approach the
%engine consisting of more than $2$ particles would always have
%negative net work gained in low temperature limit. Work could be
%gained only if all particles would be found on one of the sides of
%the cylinder, resulting in extractable work in the order of $kT$
%with probability diminishing with large $N$. On the other hand,
%all other measurement results would lead to loss of energy in the
%order of $\Delta E$ via tunnelling during the removal of the
%barrier phase.

{\em \bf Coarse-grained measurements.---} One may consider another
way of exploiting particle statistics to try to violate the second
law, by doing a coarse-grained measurement.  Let us define a
coarse-grained measurement with $M$ outcomes labelled by $m$
(running from $1$ to $M$), each occurring with probability
$p_{m}$. As $W1$ does not depend on the measurement it will not
change. The partition sum for bosons and distinguishable particles
after the measurement with outcome $m$ will have the form
$Z_{m}^{A}=\frac {1}{p_{m}}\exp\left(-NE_{1}^{r}/kT\right).$ After
the removal phase the partition sum will be
$Z_{m}^{B}=\exp\left(-NE_{1}/kT\right).$ The extractable work will
thus
be%
\begin{equation}
W=-kT\sum_{m=1}^{M}p_{m}\ln\left(  p_{m}\right) , \label{CG_work}%
\end{equation}
which holds also for fermions with $M=2$ and $p_{1}=p_{2}=1/2$.
From this we see that also in this case the extractable work does
not depend on the actual number and type of particles in the
system, but only on the measurement and its possible outcomes
(taking into account the fact that possible measurements are
limited by the number and type of particles in the system). For a
fixed number of outcomes $M$ the extractable work is maximized for
a measurement with equal probability of each outcome and will gain
$W_{\max}=kT\ln(M)$.

We note that the result in Eq. (\ref{CG_work}) can we rewritten as $W=TS$
with
\begin{equation}
S=-k\sum_{m=1}^{M}p_{m}\ln\left(  p_{m}\right) , \label{CG_entropy}%
\end{equation}
being the entropy of the measurement outcomes. This definition
exactly corresponds to the definition of Gibbs entropy except that
here one deals with the probabilities of measurement outcomes
rather than with the probabilities of microstates. One might view
this correspondence in the following way: by removing entropy $S$
from the system by a measurement one is able to extract exactly
$W=TS$ of work from the system before it turns back to its
original state. The extracted entropy has to be stored somewhere,
or erased, costing exactly the same amount of work
again~\cite{Bennett82}. (A net work yield can nevertheless be
obtained if two reservoirs with different temperatures exist and
the erasure is performed whilst in contact with the colder one.)

For a measurement of distinguishable particles, where one only measures the
number of particles on each side, the extractable work will be%
\begin{equation}
W_{d}=-kT\sum_{m=0}^{N}2^{-N}\binom{N}{m}\ln\left(  2^{-N}\binom{N}{m}\right)
, \label{Number_measurement_dist_work}%
\end{equation}
which is smaller than $W_{b}$ for bosons, where the same measurement
represents a full measurement of the system (reaching $\frac{1}{2}W_{b}$ for the
limit of large $N$ \cite{BinEntropy}). This can be explained by the smaller
information value of the measurement outcome (with the same number of possible
outcomes $N+1$) for distinguishable particles in comparison to bosons -
whereas for bosons any result is equally probable, for distinguishable
particles only the results close to balanced are likely.

{\em \bf Degeneracy.---} Let us briefly discuss the possible
degeneracies of energy levels in the system. For bosons and
distinguishable particles they only play a role for the lowest
energy level. If the degeneracy can be revealed by the
measurement, the system will correspond to an advanced SZE with
more than one barrier and more complicated measurement. If the
degeneracy cannot be revealed, influence will be exactly cancelled
for the extractable work, however it would increase the work
needed to insert the barrier.

For fermions, if the Fermi energy level of the system is degenerate, more than
one fermion can occupy the same energy. This can increase the number of
particles actually performing work in the system up to $N$. In this scenario
fermions behave exactly like bosons, condensing into the lowest energy level
and performing the same amount of work as bosons (if the degeneracies cannot
be revealed by measurement) or distinguishable particles (if degeneracies can
be revealed by the measurement).

{\em \bf Summary and discussion.---}We performed a detailed
analysis of a Szilard engine in the low temperature limit where
the working medium is either distinguishable particles, bosons or
fermions. We showed that the extractable work is determined by the
information gain of the measurement performed on the system,
regardless of the working medium. We demonstrated in detail how
things conspire to remove the dependence on the working medium in
that sense. This latter contribution is arguably of the same type
as Bennett's exorcism of Maxwell's daemon~\cite{Bennett82} in that
it shows in what exact way the second law is not violated in this
process.

%Entropy associated with such a measurement can be defined equivalently to Gibbs entropy.

We also showed that if a full measurement on the system is performed (which is associated with a different amount of information gain in the different cases), distinguishable particles
exhibit a much larger potential to deliver work (scaling linearly with the
number of particles) relative to bosons (where it scales logarithmically).
Fermions can only provide a fixed amount of work
independent of the number of particles.

Moreover we showed that for coarse-grained measurements the
extractable work is again determined by the information gain of
the measurement. This clarifies why the same kind of measurement
(measuring the number of particles on either side of the piston)
extracts more energy for bosons than for distinguishable
particles.

It would be interesting to provide experimental evidence for the
results obtained, especially for the possibility to extract work
with balanced measurement outcomes. One could think about cold
atoms kept at a stable temperature by a (larger amount) of
different atoms, as suggested in \cite{KSLU}. Another option would
be to use photons in micro-cavities as the working media. Here the
confinement potential is easily controllable and the barrier could
consist just from an inserted mirror.

{\em \bf Acknowledgements.---}We acknowledge support from the
National Research Foundation (Singapore) and the Ministry of
Education (Singapore). MP acknowledges the support of SoMoPro
project funded under FP7 (People) Grant Agreement no. 229603 and
by South Moravian Region, as well as GA\v{C}R P202/12/1142, CE SAS
QUTE and VEGA 2/0072/12.

%==============================================================================

\end{document}